\newcommand{\ls}{LS}

\newcommand{\BeNeutrino}{$^{7}$Be}
\newcommand{\BeSolarNeutrino}{$^{7}$Be solar $\rm \nu_{e}$}

\newcommand{\ThPb}{$^{212}$Pb}
\newcommand{\UPb}{$^{214}$Pb}
\newcommand{\ThRn}{$^{220}$Rn}
\newcommand{\URn}{$^{222}$Rn}
\newcommand{\Pb}{$^{210}$Pb}
\newcommand{\Lead}{Lead}
\newcommand{\Radon}{Radon}

\documentclass[
      ,final            
]
{aipproc}

\layoutstyle{8x11single}

\begin{document}

\title{Laboratory Studies of Lead Removal from Liquid Scintillator in Preparation for KamLAND's Low Background Phase}



\author{Gregory Keefer}{
  address={Lawrence Livermore National Laboratory, Livermore, California 94550, USA}
}

\begin{abstract}
The removal of \Radon\ induced \Lead\ from liquid scintillator was extensively studied in preparation for KamLAND's low background phase.
This work presents the results from laboratory experiments performed at the University of Alabama
and their implications for KamLAND and future low background experiments using carbon based liquid scintillator.  
It was observed that distillation was the most effective purification procedure and that one must consider a non-polar 
and non-ionic component of \Lead\ in order to reach the levels of radio-purity required for these new class of ultra-low 
background experiments.
\end{abstract}

\maketitle


\section{Introduction}

\par Removal of intrinsic radio-isotopes from liquid scintillator (\ls) was extensively studied in preparation 
for a low-background phase of KamLAND~\cite{bib:KeeferThesis,bib:Kishimoto2008}.  
The primary focus of these studies was to determine how to remove \Pb\ from the KamLAND \ls.  If a large enough \Pb\ reduction 
factor could be achieved, then this would be one of the primary components needed to provide KamLAND with the sensitivity to measure
 \BeSolarNeutrino.  However, \Pb\ was not the only intrinsic radioactive background that was inhibiting this measurement.
Table~\ref{tab:KamLANDConc} shows the status of the intrinsic radioactivity which was inhibiting a  \BeSolarNeutrino\ 
measurement.  A lower \Pb\ background would also reduce the correlated reactor~\cite{bib:KamLAND_PRL_3} and Geo-Neutrino 
backgrounds~\cite{bib:KamLAND-Nature}.  This is done by reducing the primary correlated background, $\rm^{12}C(\alpha,n)^{16}O$.  
Thus, a detailed program aimed at understanding the innate nature of \Pb\ removal from organic \ls\ was an important undertaking for the KamLAND experiment.

\par All future low energy solar neutrino experiments using carbon based \ls\ are subject to 
very stringent intrinsic radioactive contaminant requirements.  Background studies from KamLAND data indicated
concentration of $\rm^{210}Pb$ needed to be reduced by a factor of $10^{5}$, down to an unprecedented level of $10^{-25}$ g/g, as indicated 
in Table~\ref{tab:KamLANDConc}  This level of radio-purity equates to less than 1 atom per kg of \ls!  
It thus becomes increasingly difficult to perform measurements in the laboratory on such ultra-pure \ls.

\par This paper reports results obtained from laboratory experiments designed to target the 
purification of \ThPb\ and \UPb\ from a \ls\ mixture containing 80.2\% n-Dodecane, 19.8\% 
1,2,4-Trimethylbenzene (PC) and $1.36 \pm 0.03$ g/l of 2,5-Diphenlyoxazole (PPO) n-Dodecane, PC.
The primary methods studied for \Lead\ removal were adsorption, distillation, heating and chemical extraction.
Many other methods were also tried but did not give encouraging results and include: water extraction, 
pH dependence of water extraction, isotope exchange and filtering.

\begin{table}[h!]
  \begin{tabular}{ l l  }
      \hline
      \textbf {Isotope}  &  \textbf {Concentrations [g/g]}       \\ 
      \hline \hline
      $\rm ^{14}C $          & $\rm (3.98 \pm 0.94)\times10^{-18} $           \\ 
      $\rm ^{39}Ar $         & $\rm < 4.3 \times10^{-21}        $             \\ 
      $\rm ^{40}K $          & $\rm (1.30 \pm 0.11)\times10^{-16} $           \\ 
      $\rm ^{85}Kr $         & $\rm (6.10 \pm 0.14)\times10^{-20} $           \\ 
      $\rm ^{210}Pb$         & $\rm (2.06\pm0.04)\times10^{-20}   $           \\ 
      $\rm ^{232}Th$         & $\rm (8.24 \pm 0.49)\times10^{-17} $           \\ 
      $\rm ^{238}U$          & $\rm (1.87 \pm 0.10)\times10^{-18} $           \\ 
      \hline
    \end{tabular}
  \caption{Measured radioactivity concentrations in KamLAND~\ls.  The limit for $^{39}$Ar was derived from solubility arguments~\cite{bib:KeeferThesis}. }
  \label{tab:KamLANDConc}
\end{table}

\section{Experimental Procedure}

\par The experimental studies detailed in this paper were designed to measure concentrations of \ThPb\ 
in the laboratory down to $10^{-22}$ g/g. Measurements at this concentration are not achievable 
with \Pb\ in the laboratory.  However, it was deemed necessary to forgo using \Pb\
in exchange for understanding the underlying properties of the nucleus in \ls\ at this low level
of concentration.  At a concentration of $10^{-20}$ g/g the specific \ThPb\ activity is 0.83 Bq/l, 
a factor $1.8 \times 10^{4}$ larger than that of \Pb\ at equal concentration.  

\par The primary assumption used to compare purification techniques performed with \ThPb\ and  \UPb\ as 
apposed to using \Pb\ was that all isotopes of Lead are borne in the same manner, via Radon decay.  It was
assumed that this method of introducing Lead into the \ls\ was the key component in how the \Pb\ was actually
dispersed throughout the KamLAND \ls.

\par As such, two commercially available Radon sources were purchased from Pylon Electronics.  These Radon
sources dispersed Radon into a glass bubbler fitted with a bubbling stone and filled with \ls.  
The Radon was dispersed into the \ls\ via a filtered Nitrogen carrier gas.  The Radon decayed while inside the \ls\
and the \ThPb\ was counted on a Germanium (Ge) detector.  There were several gamma lines and nuclei that were used
to determine the initial concentration of \ThPb\ after bubbling.  The \ThPb\ loaded \ls\ was then transferred to the 
appropriate purification system and the empty Nalgene bottle, used for counting the initial \ls, was again counted on the Ge detector to 
determine the residual \ThPb\ which adhered to the bottle walls~\footnote{This was actually a significant amount of activity.  
There was no correlation with initial activity observed and the absolute amount could be as much as half the 
initial activity.  Thus, it was imperative to count the empty bottle to account for all the initial activity.}.  After the 
 \ThPb\ loaded \ls\ was purified, it was counted on the Ge detector to determine the final activity.  If needed, the \ls\ was filtered prior to counting
 to remove any particulates from the purification process.  In all instances, the procedural systematics were measured and
taken into account.  The purification factor is defined throught this work as the ratio of initial to final activity in the \ls.

\par  While this procedure works great for  \ThPb\ loaded \ls\ it is not exactly the same for \UPb\ loaded \ls\ as \URn\ is the nuclei with the largest half-life.  In order to understand the effects of \UPb\ purification we needed to take into 
account the growth and decay properly.  However, the procedure outlined above did not change, only the analysis of the data.  For 
purification procedures which achieved large reduction factors gamma counting was not sufficient.  Thus, the fast 
beta-gamma coincidence which exists in both the \URn\ and \ThRn\ chains  allowed us to achieve two orders of 
magnitude greater sensitivity in our studies, down to 10 mBq/l.  Using these two methods of counting, and the different isotopes
of \Lead, we were able to observe purification of \ThPb\ and \UPb\ in \ls\ at concentrations of $10^{-17}-10^{-22}$ g/g!

\section{Experimental Results}

\par The first experiments performed involved water.  This was done because a water extraction system already existed in the Kamioka
mine (where the KamLAND experiment exists) and thus it would mean a minimal investment if we could develop a system involving water 
extraction to achieve our desired goals.  
Unfortunately water was not ideal.  A reduction factor of $1.02 - 1.1$ was achieved with de-ionized (DI) water.  The primary method was a re-circulation
scheme in which the water was continuously circulated through the \ls.  The DI water was broken up into small bubbles before it entered 
the \ls\ to provide a greater surface to volume ratio.  To further the efforts of water, experiments were also performed which
looked at the dependence on the \ThPb\ removal relative to the pH.  There was an observed dependence  in that the the removal efficiency was higher with lower pH. However, the overall increase with a lower pH was only 15\% relative to DI water and thus was deemed ineffective.

\par  Several other types of experiments were performed that did not yield great results.  \ls\ was passed over a bed of granular \Lead\ 
shot and \Lead\ granules under the assumption that given an infinite number of stable $^{206}$Pb atoms, there would occur an isotope 
exchange of the \ThPb\ in the \ls.  However, there was no effect observed beyond that expected for the purification system itself.

\par  Filtering was a primary means of removing particulates from liquids and was used in the course of our studies.  It was observed that filtering
did purify the \ThPb\ from the \ls\ but only a factor 1.1 reduction (10\%) was observed on a single pass.  The largest reduction factors were observed
in filtering when the carrier gas used to load the \ls\ with \ThPb\ was not pre-filtered.  This resulted in factors of 1.5 reduction and can be attributed
to particulates in the gas stream ``catching'' the ionized \ThPb\ atom which is then easily removed via filtration. Once a pre-filter was placed on
the Nitrogen carrier gas, the reduction factor from filtering was consistent.  While multiple filtrations increased the reduction factor modestly it was 
not to the orders of magnitude needed for KamLAND.

\par  Adsorption was the most well studied purification technique.  This is because it was very reproducible, it yielded reasonably good purification 
results and it allowed us to look at more than just \Lead\ reduction.  Adsorption was performed with Silica Gel provided by Selecto Scientific.  
There were also experiment performed with other adsorption materials such as Cu/Mn and Alusil.  The results from these experiments are listed in Table ~\ref{Ta:AdsorbentEff}.  The details can be found in reference ~\cite{bib:KeeferThesis}.  The primary observation from the silica gel experiments was
that there was consistently approximately 5\% of the \ThPb\ which could not be removed via silica gel extraction.  The amount of silica, the type, the procedure, 
or any other physical observable could improve the purification beyond a factor 30.  Using silica gel it was also possible to obtain reduction
factors for $^{222}$Rn,  $\rm^{218}Po$,  $\rm^{212}Bi$, $\rm^{214}Bi$ and $^{212}$Bi.  These details can also be found in reference ~\cite{bib:KeeferThesis}.

\renewcommand\arraystretch{1.25}
\begin{table}[t!]
  \begin{tabular*}{\columnwidth}{@{\extracolsep{\fill}} p{3.0in} p{1.5in}@{} p{1.5in}@{}} 
      \hline
      \textbf {Adsorbent Type}  &  \textbf {\ThPb\ Reduction Factor} & \textbf {Adsorbent Mass [g]}   \\ \hline \hline
      Selecto, Lot \#301286301, Si-gel 32-63 $\mu$m       & $ 19.4 \pm 0.47 $                      &  $ 1.5 $  \\  
      Selecto, Lot \#306279301, Si-gel 100-200 $\mu$m     & $ 15.38 \pm 0.24 $                     &  $ 1.5 $  \\  
      Selecto, Lot \#102085402, Alusil 70                 & $ 27.03 \pm 0.73 $                     &  $ 1.5 $  \\ 
      Selecto, Lot \#109110402, Alusil Plus               & $ 8.33 \pm 0.07 $                      &  $ 1.5 $  \\  
      Selecto, Lot \#108110403, Alusil NanoSmart          & $ 8.20 \pm 0.07 $                      &  $ 1.5 $  \\ 
      Selecto, Lot \#900110401, Si-gel NanoSmart ACT      & $ 3.58 \pm 0.01 $                      &  $ 1.5 $  \\ 
      Selecto, Lot \#900110401, Alusil Coarse             & $ 28.57 \pm 0.82 $                     &  $ 1.5 $  \\  
      Selecto, Lot \#107223405, Alusil 40 without K       & $\rm 10.31 \pm 0.21 $                  &  $ 2.0  $    \\ 
      \raggedright Aldrich, 3-(Diethylenetriamino) Propyl-Functionalized gel          & $\rm 8.33 \pm 0.69 $       &  $ 10.0 $   \\
      \raggedright Aldrich, Triamine Tetraacetate-Functionalized gel                  & $\rm 11.11 \pm 0.12 $      &  $ 10.2 $   \\
      Aerosil  200                                         & $\rm 8.26 \pm 0.07 $                      &  $ 15.0  $   \\
      S\"{U}D-CHEMIE,  Cu/Mn Catalyst T-2550              & $\rm 3.45 \pm 0.02 $                      &  $ 0.5  $    \\ 
      S\"{U}D-CHEMIE,  Cu/Mn Catalyst T-2550, Crushed     & $\rm 26.32 \pm 0.69 $                     &  $ 7.2 $ \\ 
      $\rm Ca_{3}(PO_{4})_{2}$                            & $\rm 6.54 \pm 6.54 $                      &  $ 2.0  $    \\ 
      \hline
    \end{tabular*}
    \caption{Measured \ThPb\ reduction factors in \ls.  Only statistical errors are quoted.  Experimental investigation of the systematic error for 32-64 $\mu$m gel yielded 7.2\%. The error on the mass measurements was 0.05 g.}
    \label{Ta:AdsorbentEff}
\end{table}
\renewcommand\arraystretch{1.0}

\par The adsorption experiments presented a problem in that, adsorption addresses an ionic or polar form of a molecule or atom.  If the \Lead\ was in 
the \ls\ in a non-charged state, then adsorption and many other experiments addressing this species, such as water extraction, 
would never achieve reduction factors of $10^{5}$.  Thus it was hypothesized that the \ThPb\ as in the \ls\ in an organic form.  To test this theory, 
chemical analysis was used to specifically attack an organic \Lead\ molecule.   It was found that using FeCl$_{3}$, SnCl$_{3}$, MoS$_{2}$ and Thiol Resin
all gave increased results relative to silica gel.  SiO$_{2}$ + FeCl$_{3}$ + SiO$_{2}$ was then performed in series and a factor 1250 
reduction in \ThPb\ was achieved.  The first SiO$_{2}$ extraction was performed to remove the ``known'' amount of ionic/polar \ThPb\ from the \ls.  
Then the chemical extraction was performed
to address the non-polar form of \ThPb.  The final SiO$_{2}$ extraction was to clean up in a sense, to remove any possibly broken organic bonds that were now
polar.  While all of these chemical treatments worked, they were destructive to the \ls.  The optics of the \ls\ were completely destroyed.  However, it did
hint to a new component of \Lead\ not addressed in typical purification techniques.

\par  Another method used to test the organo-metallic \Lead\ hypothesis was heating.  Tetraethyllead actually decomposes at 200 $^{\circ}$C and thus heating
\ls\ composed of Tetraethyllead would cause the lead to decompose into a polar form which could then be extracted using silica gel.  Experiments
performed using heating in series with silica gel extraction showed that the purification factor was boosted by a factor of 10.  The total \ThPb\ reduction
found in heating plus silica gel extraction was $278 \pm 23$. Details can be found in reference ~\cite{bib:KeeferThesis}.

\par  The final purification technique addressed was distillation.  Distillation actually takes care of several of the issues discussed above without knowing 
it a priori.  The heating of the \ls\ in the distillation flask will decompose any volatile organo-metallic components of \Lead.  The distillation process itself works
under the principle of separation of states.  It was found necessary to remove the first amount of distilled liquid, the low boiling point liquid, and discard
it to achieve the best performance.  Ideally this would contain water and all the low boiling point organo-metallic \Lead\ components.
 It was further found not to distill all the distillate from the flask.  The packing material used in the distillation column acts as an adsorption material
and removes the charged \Lead\ component.  Thus, distillation was found to be the most effective.  A single pass distillation was found to remove a factor $4\times10^{3}$.  One has to be careful not to rapidly boil-over the \ls\ as this will carry impure \ls.  Typical distillation speeds in the lab were 
on the order of 10 ml/min.  Multiple distillations succeeded in achieving measured reduction factor of $10^{4}$.  The major downside to distillation is that
it is very hard to reproduce.  Measurements were performed on different apparatuses and even on the same apparatus and purification factors could vary by a factor of 10.

\section{Conclusion}

\par These laboratory studies contributed to a major paradigm shift in how we viewed the \Pb\ was dissolved in the \ls.
Experimental evidence indicated the \Lead\ impurities were present in the \ls\ in  an ionic and an organo-metallic form. 
Experiments designed to address separately the ionic form and the organic form of \Pb\ indicated that approximately 3-5\% of \Lead\ 
was present in an organo-metallic state in the \ls.  Thus, in order to achieve the substantial radio-purity needed in the
next generation experiments using organic based materials, these radioactive organo-metallic isotopes must be addressed.

\par The methods found most effective for purifying \ls\ and its components were distillation and adsorption with a combination of heating.
Distillation yielded the largest independent reduction factor for all isotopes studied, without destroying the optical properties of the \ls.  
Thus, distillation was the method used to remove \Pb\ from KamLAND \ls\ and resulted in ``The First Observation of \BeNeutrino\ 
Solar Neutrinos with KamLAND''~\cite{bib:KeeferThesis}.


\begin{theacknowledgments}
  The majority of this work was performed under the direction of my advisor, Andreas Piepke, at The University
of Alabama.  The Tohoku Research Center for Neutrino Science also participated in performing independent analysis of many of these experiments. 
\end{theacknowledgments}


\bibliographystyle{aipproc}   

\bibliography{master}

\end{document}